\documentclass[useAMS,usenatbib,usegraphicx]{mn2e}

\DeclareMathAlphabet{\mathsc}{OT1}{cmr}{m}{sc}
\def\testbx{bx}%
\DeclareRobustCommand{\ion}[2]{%
\relax\ifmmode
\ifx\testbx\f@series
{\mathbf{#1\,\mathsc{#2}}}\else
{\mathrm{#1\,\mathsc{#2}}}\fi
\else\textup{#1\,{\mdseries\textsc{#2}}}%
\fi}
\def\lesssim{\mathrel{\hbox{\rlap{\hbox{\lower5pt\hbox{$\sim$}}}\hbox{$<$}}}}
\def\gtrsim{\mathrel{\hbox{\rlap{\hbox{\lower5pt\hbox{$\sim$}}}\hbox{$>$}}}}

\title[Triggering optical AGN]{Triggering optical AGN: the need 
for cold gas, and the indirect roles of galaxy environment and 
interactions}
\author[J. Sabater, P.~N. Best and T.~M. Heckman]{J. 
Sabater$^{1}$\thanks{E-mail: jsm@roe.ac.uk (JS); pnb@roe.ac.uk (PNB)},
P.~N. Best$^{1}$
and
T.~M. Heckman$^{2}$
\\
$^{1}$Institute for Astronomy (IfA), University of Edinburgh, Royal Observatory,
Blackford Hill, EH9 3HJ Edinburgh, U.K.\\
$^{2}$Center for Astrophysical Sciences, Department of Physics \&
Astronomy, The Johns Hopkins University, Baltimore, MD 21218, USA
}

\begin{document}

\date{Accepted XXXX Month XX. Received XXXX Month XX; in original form
XXXX Month XX}

\pagerange{\pageref{firstpage}--\pageref{lastpage}} \pubyear{2014}

\maketitle

\label{firstpage}

\begin{abstract}
We present a study of the prevalence and luminosity of Active Galactic Nuclei
(AGN; traced by optical spectra) as a function of both environment and galaxy
interactions. For this study we used a sample of more than 250000 galaxies drawn
from the Sloan Digital Sky Survey and, crucially, we controlled for the effect
of both stellar mass and central star formation activity. Once these two factors
are taken into account, the effect of the local density of galaxies and of
one-on-one interactions is minimal in both the prevalence of AGN activity and
AGN luminosity. This suggests that the level of nuclear activity depends
primarily on the availability of cold gas in the nuclear regions of galaxies and
that secular processes can drive the AGN activity in the majority of cases.
Large scale environment and galaxy interactions only affect AGN activity in an
indirect manner, by influencing the central gas supply.
\end{abstract}


\section{Introduction}
\defcitealias{Sabater2013}{SBA13}

Active Galactic Nuclei (AGN) are closely linked to galaxy evolution and may play
a fundamental role in the feedback processes that both quench the growth of
massive galaxies \citep[see reviews by][and references
therein]{Cattaneo2009,Heckman2014} and establish the strong correlations between
a galaxy's black hole mass and the velocity dispersion of its stellar bulge
\citep[e.g.][]{Marconi2003,Ferrarese2000,Gebhardt2000}. The presence and
characteristics of an AGN are tightly related to those of their parent galaxies,
and nuclear activity has also been found to be linked to galaxy environment in
several studies.  Nuclear star formation (SF) activity is enhanced by
interactions \citep[see ][and references therein]{Li2008a} as is expected from
numerical simulations, but the dependence of AGN activity on environment is not
so clear. Many apparently contradictory results are still found in the
literature \citep[see discussion in][hereafter
\citetalias{Sabater2013}]{Sabater2013}.

An important point to consider is that the word ``environment'' is used to refer
to at least two relevant but distinct aspects of galaxy environment: (a) the
large scale environment, which effectively controls the gas supply to galaxies
and, (b) one-on-one interactions, involving strong tidal interactions between
companion galaxies. The relation of the prevalence of AGN with these two aspects
can be different or even opposite. For example, while the prevalence of
radiatively efficient AGN decreases toward higher local densities of galaxies
\citep{Carter2001,Miller2003,Kauffmann2004}, it is enhanced when one-on-one
interactions are stronger \citep[][]{Petrosian1982,Koulouridis2006,
Alonso2007,Rogers2009,Ellison2011,Liu2012,Hwang2012}. Hence, those different
aspects of the environment should be considered separately
\citepalias{Sabater2013}.

One of the most (if not the most) important factors that affects the triggering
of an AGN is the galaxy mass. The prevalence of AGN depend strongly on the
stellar (or black hole) mass of the host galaxy
\citep[e.g.][]{Kauffmann2003,Best2005b,Silverman2009,Tasse2011}. The relation of
the galaxy mass with the fraction of AGN is so strong that it will seriously
bias any studies that do not take it into account properly. In
\citetalias{Sabater2013}, galaxy mass was accounted for using stratified
statistical methods, and it was found that (at fixed mass) the prevalence of
optical AGN is a factor 2--3 lower in the densest environments, but increases by
a factor of $\sim 2$ in the presence of strong one-on-one interactions. The
prevalence of AGN also depends on galaxy morphology
\citep{Moles1995,Schawinski2010,Sabater2012,Hwang2012} which is closely related
to the star formation \citep{Lintott2008}. In fact, \citet{Kauffmann2007} and
\citet{LaMassa2013} show how AGN activity is better linked to the central star
formation of the galaxy than to the general star formation. \citet{Li2008} found
that, if the central star formation and the AGN prevalence are considered
together, there is no enhancement of the nuclear activity in galaxies with close
companions. Similarly, \citet{Reichard2009} found that the lopsidedness of a
galaxy is related to an enhanced activity level of the central black hole but,
if the age of the central stellar population is matched in the comparison, this
enhancement is no longer visible. These findings would be related to the
suggestion of \citet{Park2009} that galaxy mass and morphology are the main
factors determining the rest of the properties of galaxies. All of this may
indicate that the observed environmental dependence of both AGN and central star
formation arises from the same underlying mechanisms and that the environment
affects the AGN activity only by affecting the gas supply. If this is correct,
the trends of AGN fraction with large scale environment should also disappear
when the central star formation rate is controlled together with galaxy mass.
The goal of this study is to extend the analysis of \citetalias{Sabater2013} to
test this hypothesis, and also to look independently at interactions.

AGN come in two flavours depending on their feeding mechanism \citep[see][and
references therein]{Heckman2014}: (a) quasar or radiative mode, believed to be
fuelled by cold gas, observed as X-ray AGN, optical AGN and high excitation
radio galaxies, and, (b) radiatively inefficient or jet-mode, probably fuelled
by gas cooling from hot halos, and observed primarily as low excitation radio
galaxies. The relation of these two types of AGN with the environment and
interactions can be different or even opposite, as shown in
\citetalias{Sabater2013}. We will focus in this study on radiatively efficient
AGN which are supposed to be fuelled by the cold gas that can also trigger the
central star formation that we aim to control for.

We aim to study the effect of different aspects of the environment on
radiatively efficient nuclear activity after taking into account the effect of
both the mass and the star formation activity. We characterize the environment
for a sample of $\approx 270000$ galaxies using both a local galaxy density
parameter and a tracer of one-on-one interactions. The sample and the data used
are described in Section~\ref{sec:sample}. In Section~\ref{sec:res}, the
relations between the prevalence of optically selected AGN and the environmental
parameters is quantified and analysed. The results are discussed in
Section~\ref{sec:dis} and the final conclusions are presented in
Section~\ref{sec:con}. \label{sec:cosmo} Throughout the paper, the following
cosmological parameters are assumed: $\Omega_{\mathrm{m}}=0.3$,
$\Omega_{\mathrm{\Lambda}}=0.7$ and $H_{\mathrm{0}}=
70\,\mathrm{km}\,\mathrm{s^{-1}}\,\mathrm{Mpc^{-1}}$.


\section{The sample and the data}
\label{sec:sample}

We use the sample presented in \citetalias{Sabater2013}. This sample was based
on the seventh data release of the Sloan Digital Sky Survey \citep[SDSS
DR7;][]{Abazajian2009}. It is composed of galaxies in the main spectroscopic
sample with magnitudes between 14.5 and 17.77 in r-band \citep{Strauss2002} and
with redshift between 0.03 and 0.1. The final number of galaxies in the sample
is 267973.\footnote{There were duplicated data for 4 galaxies in the catalogue
presented on \citetalias[$n_{\mathrm{rows}}=267977$;][]{Sabater2013}. The
indices of these galaxies are: 51930-285-80; 51999-286-559; 52173-644-540 and
52468-717-223 (mjd-plate-fiberid). The duplicate entries were removed from the
catalogue.}

We will use two of the environmental parameters derived in
\citetalias{Sabater2013}. In that study three environmental parameters where
considered, to trace different aspects of the environment and interaction: (a) a
local galaxy density parameter (hereafter ``density''), (b) a tidal forces
estimator, and (c) a cluster richness estimator \citep[from][]{Tago2010}. The
density parameter is derived from the density of galaxies up to the 10$^{th}$
nearest neighbour; defined as $\log(10/\mathrm{Vol}(r_{10^{th}}))$, where
$r_{10^{th}}$ is the projected distance in Mpc to the 10$^{th}$ companion. The
tidal estimator traces the relation between the tidal forces exerted by
companions and the internal binding forces of the galaxy; defined as
$\log(\sum_i [(L_{r}/L_{r_{i}}) \times (2R/d_{i})^{3}])$, where $L_{r}$ is the
luminosity in r-band of the galaxy, $L_{r_{i}}$ the luminosity in r-band of the
companion, $R$ the radius of the galaxy and $d_{i}$ the distance between the
galaxy and the companion. A Principal Component Analysis (PCA) was applied to
consider and remove the possible correlations between the environmental
parameters. We found that the local density of companions around the target
galaxy (a measure of the larger-scale environment in which the galaxy lives) is
one of the main environmental driving factors for AGN activity and was well
traced by the density parameter (also largely equivalent to the PCA1 component).
We found that one-on-one interactions are also important driving factors, and
are best traced by the the PCA component denominated PCA2 (defined as $0.707
\times \mathrm{tidal} - 0.707 \times \mathrm{density}$). PCA2 seems to take into
account and corrects, at least partially, the possible 2D projection effects
that would affect a pure tidal estimator in dense environments. Hence, we
selected density and PCA2 for this study.

We also used the total stellar mass of the galaxy and the specific star
formation rate measured in the central area covered by the spectrograph fibre
(\citealt{Kauffmann2003, Brinchmann2004}; hereafter we will use sSFR to refer to
this central measurement of the specific star formation rate: sSFR $=$ nuclear
star formation divided by total stellar mass). We will also use the optical AGN
classification. A galaxy is considered to harbour an optical AGN if it is
classified as a Seyfert, LINER or transition object using the standard emission
line ratio diagnostic diagrams \citep[e.g.][]{Kewley2006} and the luminosity of
its [\ion{O}{iii}] emission line is higher than $10^{6.5}\, \mathrm{L_{\odot}}$.
This limit selects only bright AGN but avoids the possible classification biases
arising from the different redshifts of the galaxies of the sample
\citepalias[see][]{Sabater2013}. The number of AGN galaxies classified as each
type using the former criteria are the following: 3347 Seyfert, 644 LINER, 1704
transition objects. The mean $L_{\mathrm{[\ion{O}{iii}]}}$ of the AGN is $6.90
\pm 0.34$. Note that because of the high $L_{\mathrm{[\ion{O}{iii}]}}$ limit,
the vast majority of the selected AGN are Seyferts, so this will be a study of
radiatively efficient AGN, even if LINERs are classified as jet-mode AGN
\citep{Heckman2014}.

There is a chance for relatively faint AGN to be misclassified as SF nuclei if
their emission is concealed by the strong emission of a powerful star forming
host. The high $L_{\mathrm{[\ion{O}{iii}]}}$ limit helps to minimise the
misclassification of AGN within galaxies with high star formation rates.
Furthermore, for galaxies of a given mass and SFR, there is no reason to expect
that any misclassification should be a function of environment. To check this,
we examined AGN-tracers (e.g. $L_{\mathrm{[\ion{O}{iii}]}}$) in the
SF-classified galaxies in both high and low density (or PCA2) environments and
found no evidence of any differences. We also confirmed that our results were
unchanged (within the errors) if transition objects (a limit case mix of SF and
AGN) were excluded from the analysis. Therefore, this effect should not affect
the results.

\begin{figure}
\centering
 \includegraphics[width=8cm]{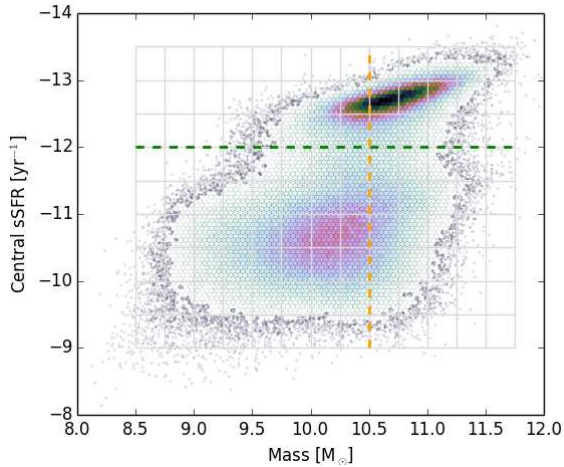}
  \caption{Distribution of mass and sSFR of the whole sample. The separation
lines of sSFR $= 10^{-12}\, \mathrm{yr^{-1}}$ and $M = 10^{10.5}\,
\mathrm{M_{\odot}}$ are shown as dashed lines. The grid marks the bins used for
some of the stratified statistical studies of Section~\ref{sec:res} (see text).}
  \label{fig:mass_ssfr}
\end{figure}

The distribution of the mass and sSFR is shown in Fig.~\ref{fig:mass_ssfr}. The
mass of the galaxies of the sample ranges from a minimum of $10^{8.1}\,
\mathrm{M_{\odot}}$ to a maximum of $10^{12.2}\, \mathrm{M_{\odot}}$ with a mean
of $10^{10.3}\, \mathrm{M_{\odot}}$. The sSFR follows the well-established
distribution \citep[e.g.][]{Strateva2001} with a star forming population (sSFR
$\gtrsim 10^{-12}\, \mathrm{yr^{-1}}$) and a passive population (sSFR $\lesssim
10^{-12}\, \mathrm{yr^{-1}}$) and 99 per cent of the galaxies with values
between $10^{-9.4}\, \mathrm{yr^{-1}}$ and $10^{-13.2}\, \mathrm{yr^{-1}}$. We
will use a value of sSFR$ = 10^{-12}\, \mathrm{yr^{-1}}$ to separate the high
and low sSFR populations when needed. An additional separation in mass of $M =
10^{10.5}\, \mathrm{M_{\odot}}$ will be used as well when required.

\begin{figure*}
\centering
 \includegraphics[width=10cm]{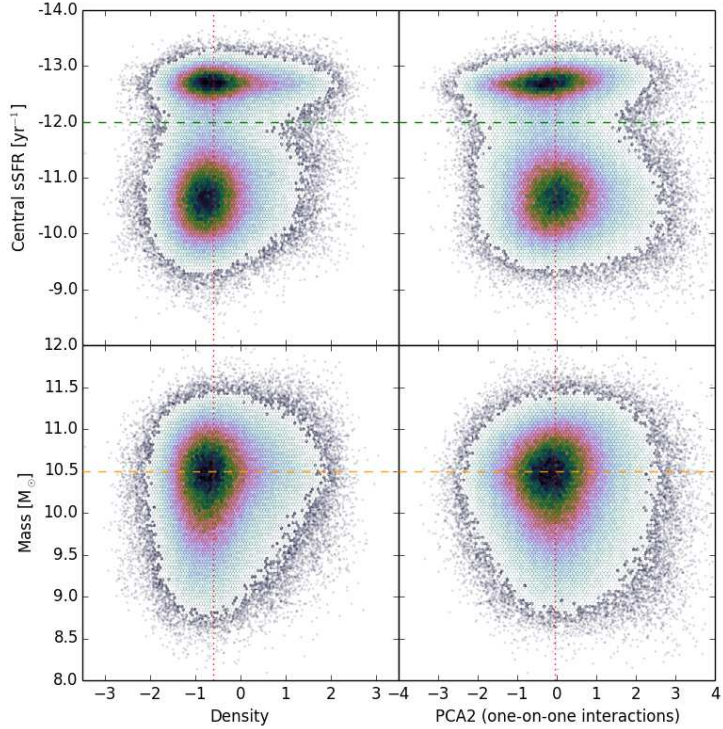}
  \caption{Distribution of mass, sSFR, density and PCA2 for the whole sample.
The medians of the density and PCA2 are marked with dotted lines. The separation
lines of $\mathrm{sSFR} = 10^{-12}\, \mathrm{yr^{-1}}$ and $M = 10^{10.5}\,
\mathrm{M_{\odot}}$ are shown as dashed lines.}
  \label{fig:params}
\end{figure*}

The relation of the environmental parameters, density and PCA2, with respect to
the mass and the sSFR is shown in Fig.~\ref{fig:params}. The correlations are
subtle but may be strong enough to bias the study if not taken into account at a
later stage. The two sSFR populations are clearly visible on the upper panels.
The high density end is populated by massive galaxies (visible on the lower-left
panel) and these galaxies are mainly low sSFR galaxies (upper-left panel). There
is a weak trend for low sSFR galaxies to be located at lower levels of PCA2 than
high sSFR galaxies (upper-right panel).

Finally, for some computations we used the mass of the black hole, which was
derived using the relation of \citet{McConnell2013} as explained in
\citet{Heckman2014}. We also considered the bolometric luminosity of the AGN to
be $L_{\mathrm{bol}} = 3.5\times10^{3} L_{\mathrm{[\ion{O}{iii}]}}$, where
$L_{\mathrm{[\ion{O}{iii}]}}$ is not corrected for dust extinction.


\section{Results}
\label{sec:res}

We investigate the relation of the prevalence of optical AGN with the
environmental parameters, density and PCA2, taking into account the effect of
the mass and the sSFR. To do that, we compute the odds ratio of a galaxy
harbouring an AGN at values of the environmental parameter above or below its
median value ($-0.603$ for the density and $-0.043$ for PCA2; see
Fig.~\ref{fig:params}). The masses are divided in strata of 0.25 [M$_\odot$]
from $\log(M) = 8.50 \, \mathrm{[M_{\odot}]}$ to $\log(M) = 11.75 \,
\mathrm{[M_{\odot}]}$, and the sSFR are divided in strata of 0.5 [yr$^{-1}$]
from $\log(\mathrm{sSFR}) = -13.5 \, \mathrm{[yr^{-1}]}$ to $\log(\mathrm{sSFR})
= -9.0 \, \mathrm{[yr^{-1}]}$. The grid composed by the different strata is
shown in Fig.~\ref{fig:mass_ssfr}. In each of these bins, a Fisher's exact test
is performed. We obtain the odds ratio (the strength of the relation between
harbouring an AGN and a higher value of the environmental parameter; if this is
$\approx 1$ the chance of harbouring an optical AGN is the same at lower and
higher values of the environmental parameter) and a $p$-value that indicates
whether the trend found is significant.

\begin{figure*}
\centering
 \includegraphics[width=15cm]{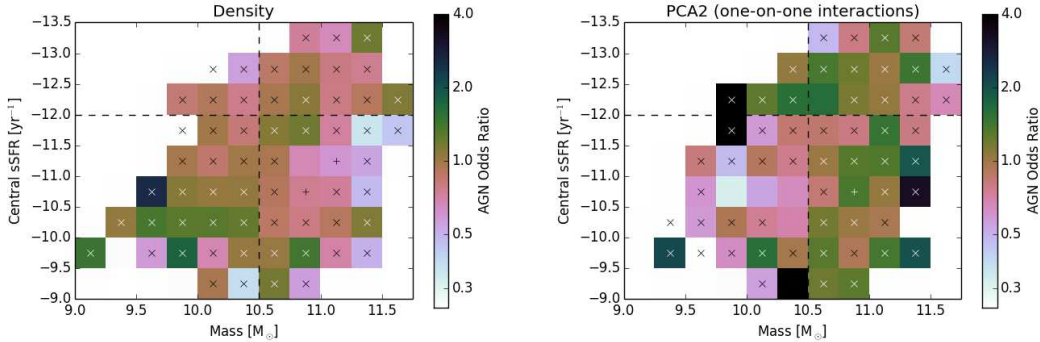}
  \caption{Odds ratios of the prevalence of AGN with respect to the level of
density and one-on-one interactions (PCA2). The bins correspond to those shown
in Fig.~\ref{fig:mass_ssfr}. An odds ratio lower than one indicates a negative
relation between the probability of harbouring an AGN and a higher value of the
environmental parameter; a value higher than one indicates a positive relation.
Note that the statistical significance must be taken into account in order to
interpret these relations. The bins for which the $p$-value is higher than 0.05
are marked with a cross and those in which the $p$-value is between 0.05 and
0.01 are marked with a plus sign. Only in the one (left panel) and six (right
panel) bins with no symbol is the odds ratio significantly ($p \leq 0.01$) at
variance with unity.}
  \label{fig:or}
\end{figure*}

The results of the Fisher's test are shown in Fig.~\ref{fig:or}. For the density
parameter, the odds ratios are usually close to 1 and the $p$-values are, in
general, higher than 0.05 indicating that the possible trends are not
significant. There is only one bin in the high sSFR and high mass region where
the trend may be significant ($p \leq 0.01$) but the value of the odds ratio is
not far from 1 in this case. We therefore find no significant trend of the
prevalence of AGN with respect to the density. In the case of PCA2, we find a
significant positive trend for one bin in the low sSFR and high mass region and
a clearly significant negative trend for 4 bins in the high sSFR and low mass
region. 

It should be noted that the size of the bins may affect the statistics of the
test. The bins need to be small to avoid biases due to strong trends of the
fraction of AGN with mass, but the smaller the bins the lower the statistical
significance of the test. Therefore, we will use a statistical method that
aggregates the information of the bins to gain in statistical significance but
that still accounts for the different strata used: the Cochran-Mantel-Haenzsel
test \citep[CMH;][]{Cochran1954,Mantel1959}. It gives the strength of
association between two bi-valued variables (odds ratio) and its statistical
significance after taking into account the effects of the possible confounding
factors defined as the strata.

It is also clear that the whole sample may not homogeneously follow a trend;
that appears to be the case for PCA2 (right panel of Fig.~\ref{fig:or}). A
Woolf's test of homogeneity of odds ratios among strata \citep{woolf1955} was
applied to check whether the trends were homogeneous enough within the sample. A
low $p$-value for this test means that the trends are different for different
parts of the parameter space. Hence, we need to separate the analysis in
different regions to obtain meaningful values of the CMH test. For density, we
found no evidence of any heterogeneity. In the case of PCA2, we found signs of
statistically significant heterogeneity using this test for the whole sample and
also when considering only the high sSFR sub-sample. However, when splitting
into four sub-samples, by mass and sSFR, no evidence of heterogeneity was found
in PCA2 for any sub-sample.

We applied the CMH test to check for the significance of the relation between
the presence of an optical AGN and values of the environmental parameters above
or below their median. The confounding factors considered are again sSFR and
mass. The test was applied to the complete sample, to the high and low sSFR
sub-samples and to the four sub-samples obtained with the sSFR $= 10^{-12}\,
\mathrm{yr^{-1}}$ and $M = 10^{10.5}\, \mathrm{M_{\odot}}$ separation lines. We
checked that the results were not affected by the sizes of the bins. The results
are the same using from 4 to 128 bins per sample or sub-sample.

\begin{figure*}
\centering
 \includegraphics[width=14cm]{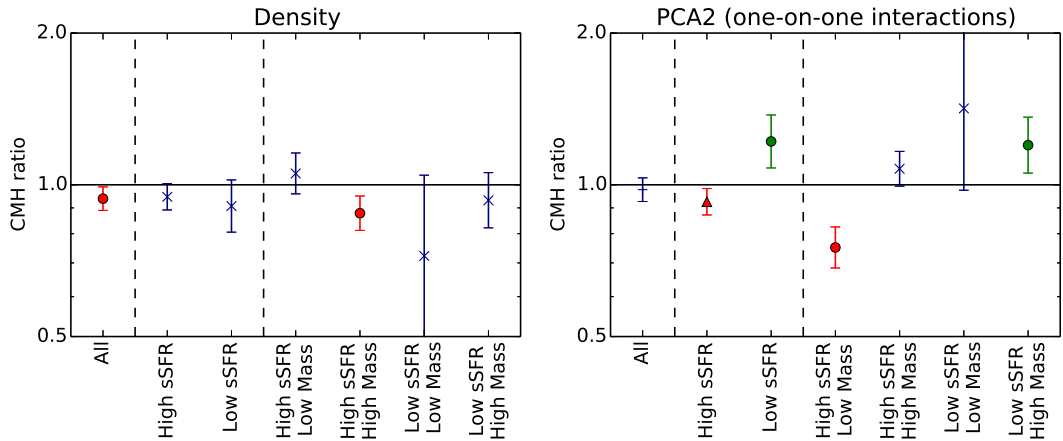}
  \caption{CMH ratios for the prevalence of optical AGN with respect to the
density and one-on-one interactions (PCA2). The error bars mark the 95 per cent
confidence interval. The shape of the symbol indicates if the trend found is
statistically significant and if the data of the sample tested is homogeneous
enough for the CMH ratio to be relevant (see text). Statistically significant
trends ($p \leq 0.05$) are marked with solid symbols (triangles or circles). A
value of the Woolf test below 0.05, indicating that the odds ratios are
heterogeneous among the strata, is marked as a plus sign (instead of a cross) or
a triangle (instead of a circle).}
  \label{fig:cmh}
\end{figure*}

The results of the CMH and the Woolf's tests are shown in
Table~\ref{table:stats} and Fig.~\ref{fig:cmh}. The values of the ratios for the
density are in general compatible with the absence of a clear trend. Even in the
two cases where the deviation of the ratios from unity can be considered
statistically significant, their values are very close to 1 implying changes in
the probability of triggering an AGN of less than 10 per cent. In the case of
PCA2 the result for the whole sample is compatible with 1 but with
heterogeneity. When the sample is separated by sSFR an opposite significant
trend appears in the two sSFR regimes; galaxies with low sSFR and higher values
of PCA2 harbour more AGN (about 20 per cent more when the value of PCA2 is above
the mean). On the other hand, galaxies with high sSFR, especially galaxies with
low mass and high sSFR seem to harbour fewer AGN in this case. The AGN
prevalence is about 30 per cent higher at low than at high PCA2 values for high
sSFR, low mass galaxies.

\begin{table}
\centering
\caption{Statistical study results from the CMH test. For each cell of the table
the first row shows the CMH common odds ratio and its 95 per cent confidence
interval and the second row shows the statistical significance or $p$-value
(i.e., the probability of the trend occurring by chance) measured by the CMH
test. The hypothesis tested was that the observed nuclear activity type is
independent of the density or PCA2 parameter. The typeface of the $p$-value
depends on its value: bold if $p < 0.01$; bold italics if $0.01 \leq p < 0.05$;
and italics if $p \geq 0.05$. An asterisk marks the two results where the data
is deemed as heterogeneous using the Woolf test.}
  \label{table:stats}
\begin{tabular}{lcc}
\hline
 & Density & PCA2 \\
\hline 
\hline
All & $0.939 \pm 0.049$ & $0.978 \pm 0.051$ * \\
 & \textit{\textbf{0.0214}} & \textit{0.4259}\\
High sSFR & $0.947 \pm 0.055$ & $0.925 \pm 0.054$ * \\
 & \textit{0.0761} & \textit{\textbf{0.0124}}\\
Low sSFR & $ 0.91 \pm  0.10$ & $ 1.22 \pm  0.14$\\
 & \textit{0.1176} & \textbf{0.0014}\\
High sSFR \& Low Mass & $1.053 \pm 0.094$ & $0.751 \pm 0.067$\\
 & \textit{0.2855} & \textbf{0.0000}\\
High sSFR \& High Mass & $0.878 \pm 0.066$ & $1.076 \pm 0.082$\\
 & \textbf{0.0013} & \textit{0.0741}\\
Low sSFR \& Low Mass & $ 0.72 \pm  0.22$ & $ 1.42 \pm  0.44$\\
 & \textit{0.1006} & \textit{0.0823}\\
Low sSFR \& High Mass & $ 0.93 \pm  0.11$ & $ 1.20 \pm  0.14$\\
 & \textit{0.2870} & \textbf{0.0058}\\
\hline
\end{tabular}
\end{table}

\begin{figure*}
\centering
 \includegraphics[width=15cm]{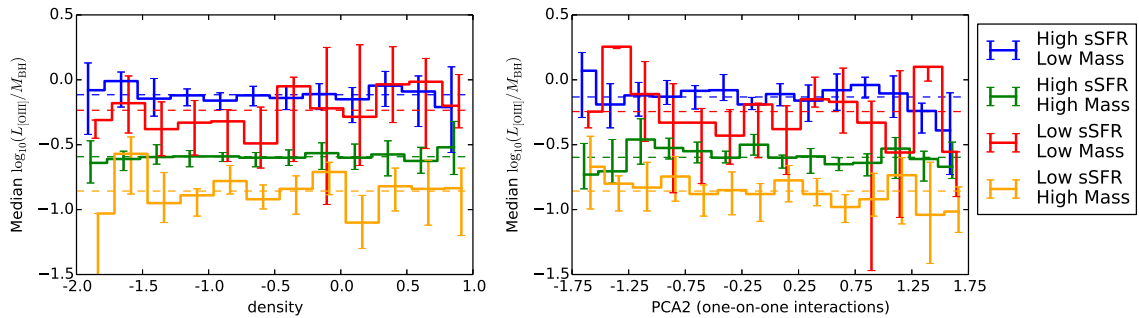}
  \caption{Activity level of the AGN traced by the median of the 
$\log(L_{\mathrm{[\ion{O}{iii}]}}/M_{\mathrm{BH}})$ with respect to the
density and one-on-one interactions (PCA2) for each of the sub-samples. 
The error bars mark the 95 per cent confidence interval. 
The dashed lines mark the median for each sub-sample.}
  \label{fig:activity_level}
\end{figure*}

We checked the activity level of the AGN with respect to the environmental
parameters. We computed the median of the logarithm of the [\ion{O}{iii}]
luminosity per unit black hole mass, which is a measure of the Eddington-scaled
accretion rate ($L_{\mathrm{bol}}/L_{\mathrm{Edd}} \sim 0.11
L_{\mathrm{[\ion{O}{iii}]}}/M_{\mathrm{BH}}$), at different levels of the
environmental parameters in the four different sub-samples. The results are
shown in Fig.~\ref{fig:activity_level}. The value of the median
$\log(L_{\mathrm{[\ion{O}{iii}]}}/M_{\mathrm{BH}})$ does not show a significant
trend with the density or PCA2 in any of the sub-samples and is, in general,
compatible with the median value for each sub-sample. On the other hand, the
median value can be seen to depend strongly on both the mass and sSFR, which are
used to define the sub-samples, emphasising the need to account for these
parameters.

Finally, we checked if any possible contamination by jet mode AGN was affecting
our results. We applied a cut-off on the accretion rate of the AGN selecting
only galaxies with $L_{\mathrm{bol}}/L_{\mathrm{Edd}} \geq 0.01$ and repeated
the study. More than 80 per cent of the galaxies were selected and the results
were the same.


\section{Discussion}
\label{sec:dis}

The test of the prevalence AGN with respect to the density is fully compatible
with the hypothesised scenario in which the galaxy density does not play any
role after the galaxy mass and the central sSFR are taken into account. The
$p$-value is 2 per cent and the CMH ratio is very close to 1. If the sample is
separated by sSFR the $p$-values are above 5 per cent. Given the size of the
sample and the magnitude of the CMH ratio, we can consider that there is not any
significant trend with respect to the density of galaxies. We showed in
\citetalias{Sabater2013} how the trend is significant if only the mass is
considered. Hence, the effect of the density and the central sSFR are strongly
correlated. Additionally, we found the lack of a clear trend for the median of
$\log(L_{\mathrm{[\ion{O}{iii}]}}/M_{\mathrm{BH}})$, which traces the activity
level of the AGN, with respect to density. The main variation of the activity
level is driven by the galaxy mass and the central sSFR. All of this is in
agreement with the picture arising from the results of \citet{Li2008} and
\citet{Reichard2009} that the AGN activity depends on the cold gas supply to the
nucleus and not on how it gets there. In higher galaxy density environments the
gas supply to the galaxy as a whole is reduced, and reduced gas supply to the
centre indirectly reduces AGN activity. 

Our own study on the effect of one-on-one interactions is also broadly
consistent with this picture. We find that the activity level of the AGN (as
judged from the median of $\log(L_{\mathrm{[\ion{O}{iii}]}}/M_{\mathrm{BH}})$)
is independent of the PCA2 parameter, once mass and sSFR are accounted for. This
result is in agreement with that of \citet{Reichard2009}, who found that the
activity level depends clearly on the stellar age \citep[traced by D4000 and
correlated with the sSFR;][]{Brinchmann2004} but with little dependence on the
lopsidedness of the galaxy (their Fig.~14). Our CMH test for the sample as a
whole is also consistent with the lack of any trend between AGN prevalence and
PCA2, in line with this picture. However, if the sample is separated into
sub-samples some significant trends arise at a 20 to 30 per cent level. Although
significant, they are small in comparison to some of the high ratio values found
in \citetalias{Sabater2013} when central sSFR was not accounted for. The trends
found do not directly fit into the previous model and suggest additional factors
may be at play. 

The increase found for low sSFR galaxies could be explained by the expected
positive relation between the triggering of AGN and strong interactions present
in theoretical models \citep[e.g.][]{Hopkins2006}. However, under this
assumption a similar trend would also be expected for high sSFR galaxies and
this is not the seen. A more likely explanation is the different time-scale
between the triggering of the central star formation burst and the AGN
\citep[e.g.][]{Li2008,Darg2010,Wild2010}. \citet{Wild2010} found that the growth
of the black hole is delayed by about 250 Myr with respect to the start of the
starburst. This time delay may be sufficient to weaken the correlation between
central SF and AGN activity in interacting systems (high PCA2) and cause the
observed trends: in interacting galaxies, AGN activity is less likely to occur
when the sSFR is high (when the starburst is triggered) and more likely to be
observed later, when the sSFR has dropped. 

In general, we find that central sSFR and galaxy density are strongly correlated
and one-on-one interactions play a secondary role on the triggering of AGN. This
results are compatible with a scenario in which the presence of cold gas in the
centre of a galaxy  is the principal factor required for the triggering of
radiatively efficient AGN. The cold gas supply to the central regions depends
upon both the galaxy's large-scale environment and any interactions. The AGN
does not need one-on-one interactions to be triggered and can be fed via secular
processes.


\section{Summary and conclusions}
\label{sec:con}

We presented a study of the relation between the prevalence of optical nuclear
activity with respect to the environment in a sample of SDSS galaxies. We aimed
to quantify the effect of different aspects of the environment on radiatively
efficient nuclear activity. The study is based on a sample of $\sim$270000
galaxies drawn from the DR7 of the SDSS. Two environmental parameters, the local
density of galaxies and PCA2 which traces one-on-one interactions, were used.
The mass and sSFR of the galaxy were considered as confounding factors of
several stratified statistical tests. The sample was divided in sub-samples
based on the mass and the sSFR when necessary and the significance of the
prevalence of AGN with respect to the environmental parameters was obtained. We
found that:

\begin{enumerate}
   \item The dependence of the prevalence of AGN on the local galaxy density is
practically null once both the mass and the sSFR are fixed. This indicates that
AGN activity depends only on the availability of cold gas at the centre of a
galaxy.
   \item The variation of the AGN activity level, traced by an Eddington-scaled
accretion rate parameter, seems to be mainly driven by the galaxy mass and the
central sSFR. Once their effect is taken into account, the effect of the density
and one-on-one interactions is secondary at most.
   \item The effect of PCA2 (one-on-one interactions) is not significant if the
sample is considered as a whole. However, a general homogeneous trend is not
found and there are significant trends when the mass and sSFR sub-samples are
considered separately: (a) a small positive trend on the prevalence of AGN with
one-on-one interactions for galaxies with high mass and low sSFR, and, (b) a
slight negative relation on the prevalence of AGN for low mass high sSFR
galaxies with respect to one-on-one interactions. The interpretation of those
trends may be related to the delay in the onset of an AGN after
the galaxy interacts and the star formation is enhanced.
\end{enumerate}

Overall, the effect of the local density of galaxies and of one-on-one
interactions is minimal in both the prevalence of AGN activity and AGN
luminosity, once the effects of mass and central star-formation are accounted
for. This suggests that the level of nuclear activity depends primarily on the
availability of cold gas in the nuclear regions of galaxies and that secular
processes can drive the AGN activity in the majority of cases. Large scale
environment and galaxy interactions only affect AGN activity in an indirect
manner, by influencing the central gas supply.

\section*{Acknowledgments}
\defcitealias{astropy}{Astropy Collaboration}
JS and PNB are grateful for financial support from STFC. We thank the referee
for quick and helpful comments.

This research made use of \textsc{Astropy}, a community-developed core Python
package for
Astronomy (\citetalias{astropy} \citeyear{astropy}); \textsc{Ipython}
\citep{IPython}; \textsc{matplotlib} \citep{matplotlib}; \textsc{numpy}
\citep{numpy}; \textsc{pandas} \citep{pandas}; \textsc{scipy}
\citep{scipy} and \textsc{TOPCAT} \citep{TOPCAT}.

Funding for the SDSS and SDSS-II has been provided by the Alfred P.
Sloan Foundation, the Participating Institutions, the National Science
Foundation, the U.S. Department of Energy, the National Aeronautics and Space
Administration, the Japanese Monbukagakusho, the Max Planck Society, and the
Higher Education Funding Council for England. The SDSS Web Site is
http://www.sdss.org/. The SDSS is managed by the Astrophysical Research
Consortium for the Participating Institutions. The Participating Institutions
are the American Museum of Natural History, Astrophysical Institute Potsdam,
University of Basel, University of Cambridge, Case Western Reserve University,
University of Chicago, Drexel University, Fermilab, the Institute for Advanced
Study, the Japan Participation Group, Johns Hopkins University, the Joint
Institute for Nuclear Astrophysics, the Kavli Institute for Particle
Astrophysics and Cosmology, the Korean Scientist Group, the Chinese Academy of
Sciences (LAMOST), Los Alamos National Laboratory, the Max-Planck-Institute for
Astronomy (MPIA), the Max-Planck-Institute for Astrophysics (MPA), New Mexico
State University, Ohio State University, University of Pittsburgh, University of
Portsmouth, Princeton University, the United States Naval Observatory, and the
University of Washington.

\bibliographystyle{mn2e_warrick}
\bibliography{databasesol}

\begin{thebibliography}{50}
\providecommand{\natexlab}[1]{#1}

\bibitem[{{Abazajian} et~al.(2009)}]{Abazajian2009}
{Abazajian} K.~N. et~al., 2009, \apjs, 182, 543

\bibitem[{{Alonso} et~al.(2007){Alonso}, {Lambas}, {Tissera} \&
  {Coldwell}}]{Alonso2007}
{Alonso} M.~S., {Lambas} D.~G., {Tissera} P., {Coldwell} G., 2007, \mnras, 375,
  1017

\bibitem[{{Astropy Collaboration} et~al.(2013)}]{astropy}
{Astropy Collaboration} et~al., 2013, \aap, 558, A33

\bibitem[{{Best} et~al.(2005){Best}, {Kauffmann}, {Heckman}, {Brinchmann},
  {Charlot}, {Ivezi{\'c}} \& {White}}]{Best2005b}
{Best} P.~N., {Kauffmann} G., {Heckman} T.~M., {Brinchmann} J., {Charlot} S.,
  {Ivezi{\'c}} {\v Z}., {White} S.~D.~M., 2005, \mnras, 362, 25

\bibitem[{{Brinchmann} et~al.(2004){Brinchmann}, {Charlot}, {White},
  {Tremonti}, {Kauffmann}, {Heckman} \& {Brinkmann}}]{Brinchmann2004}
{Brinchmann} J., {Charlot} S., {White} S.~D.~M., {Tremonti} C., {Kauffmann} G.,
  {Heckman} T., {Brinkmann} J., 2004, \mnras, 351, 1151

\bibitem[{{Carter} et~al.(2001){Carter}, {Fabricant}, {Geller}, {Kurtz} \&
  {McLean}}]{Carter2001}
{Carter} B.~J., {Fabricant} D.~G., {Geller} M.~J., {Kurtz} M.~J., {McLean} B.,
  2001, \apj, 559, 606

\bibitem[{{Cattaneo} et~al.(2009)}]{Cattaneo2009}
{Cattaneo} A. et~al., 2009, \nat, 460, 213

\bibitem[{Cochran(1954)}]{Cochran1954}
Cochran W.~G., 1954, Biometrics, 10, 417

\bibitem[{{Darg} et~al.(2010)}]{Darg2010}
{Darg} D.~W. et~al., 2010, \mnras, 401, 1552

\bibitem[{{Ellison} et~al.(2011){Ellison}, {Patton}, {Mendel} \&
  {Scudder}}]{Ellison2011}
{Ellison} S.~L., {Patton} D.~R., {Mendel} J.~T., {Scudder} J.~M., 2011, \mnras,
  418, 2043

\bibitem[{{Ferrarese} \& {Merritt}(2000)}]{Ferrarese2000}
{Ferrarese} L., {Merritt} D., 2000, \apjl, 539, L9

\bibitem[{{Gebhardt} et~al.(2000)}]{Gebhardt2000}
{Gebhardt} K. et~al., 2000, \apjl, 539, L13

\bibitem[{{Heckman} \& {Best}(2014)}]{Heckman2014}
{Heckman} T.~M., {Best} P.~N., 2014, \araa, 58

\bibitem[{{Hopkins} \& {Hernquist}(2006)}]{Hopkins2006}
{Hopkins} P.~F., {Hernquist} L., 2006, \apjs, 166, 1

\bibitem[{Hunter(2007)}]{matplotlib}
Hunter J.~D., 2007, Computing in Science \& Engineering, 9, 90

\bibitem[{{Hwang} et~al.(2012){Hwang}, {Park}, {Elbaz} \& {Choi}}]{Hwang2012}
{Hwang} H.~S., {Park} C., {Elbaz} D., {Choi} Y.~Y., 2012, \aap, 538, A15

\bibitem[{Jones et~al.(2001--)Jones, Oliphant, Peterson et~al.}]{scipy}
Jones E., Oliphant T., Peterson P. et~al., 2001--, {SciPy}: Open source
  scientific tools for {Python}. [Online; accessed 2014-08-26]

\bibitem[{{Kauffmann} et~al.(2003)}]{Kauffmann2003}
{Kauffmann} G. et~al., 2003, \mnras, 346, 1055

\bibitem[{{Kauffmann} et~al.(2004)}]{Kauffmann2004}
{Kauffmann} G., {White} S.~D.~M., {Heckman} T.~M., {M{\'e}nard} B.,
  {Brinchmann} J., {Charlot} S., {Tremonti} C., {Brinkmann} J., 2004, \mnras,
  353, 713

\bibitem[{{Kauffmann} et~al.(2007)}]{Kauffmann2007}
{Kauffmann} G. et~al., 2007, \apjs, 173, 357

\bibitem[{{Kewley} et~al.(2006){Kewley}, {Groves}, {Kauffmann} \&
  {Heckman}}]{Kewley2006}
{Kewley} L.~J., {Groves} B., {Kauffmann} G., {Heckman} T., 2006, \mnras, 372,
  961

\bibitem[{{Koulouridis} et~al.(2006){Koulouridis}, {Plionis}, {Chavushyan},
  {Dultzin-Hacyan}, {Krongold} \& {Goudis}}]{Koulouridis2006}
{Koulouridis} E., {Plionis} M., {Chavushyan} V., {Dultzin-Hacyan} D.,
  {Krongold} Y., {Goudis} C., 2006, \apj, 639, 37

\bibitem[{{LaMassa} et~al.(2013){LaMassa}, {Heckman}, {Ptak} \&
  {Urry}}]{LaMassa2013}
{LaMassa} S.~M., {Heckman} T.~M., {Ptak} A., {Urry} C.~M., 2013, \apjl, 765,
  L33

\bibitem[{{Li} et~al.(2008{\natexlab{a}}){Li}, {Kauffmann}, {Heckman}, {Jing}
  \& {White}}]{Li2008a}
{Li} C., {Kauffmann} G., {Heckman} T.~M., {Jing} Y.~P., {White} S.~D.~M.,
  2008{\natexlab{a}}, \mnras, 385, 1903

\bibitem[{{Li} et~al.(2008{\natexlab{b}}){Li}, {Kauffmann}, {Heckman}, {White}
  \& {Jing}}]{Li2008}
{Li} C., {Kauffmann} G., {Heckman} T.~M., {White} S.~D.~M., {Jing} Y.~P.,
  2008{\natexlab{b}}, \mnras, 385, 1915

\bibitem[{{Lintott} et~al.(2008)}]{Lintott2008}
{Lintott} C.~J. et~al., 2008, \mnras, 389, 1179

\bibitem[{{Liu} et~al.(2012){Liu}, {Shen} \& {Strauss}}]{Liu2012}
{Liu} X., {Shen} Y., {Strauss} M.~A., 2012, \apj, 745, 94

\bibitem[{Mantel \& Haenszel(1959)}]{Mantel1959}
Mantel N., Haenszel W., 1959, Journal of the National Cancer Institute, 22, 719

\bibitem[{{Marconi} \& {Hunt}(2003)}]{Marconi2003}
{Marconi} A., {Hunt} L.~K., 2003, \apjl, 589, L21

\bibitem[{{McConnell} \& {Ma}(2013)}]{McConnell2013}
{McConnell} N.~J., {Ma} C.~P., 2013, \apj, 764, 184

\bibitem[{McKinney(2010)}]{pandas}
McKinney W., 2010, in S.~van~der Walt, J.~Millman, eds, Proceedings of the 9th
  Python in Science Conference. pp. 51 -- 56

\bibitem[{{Miller} et~al.(2003){Miller}, {Nichol}, {G{\'o}mez}, {Hopkins} \&
  {Bernardi}}]{Miller2003}
{Miller} C.~J., {Nichol} R.~C., {G{\'o}mez} P.~L., {Hopkins} A.~M., {Bernardi}
  M., 2003, \apj, 597, 142

\bibitem[{{Moles} et~al.(1995){Moles}, {Marquez} \& {Perez}}]{Moles1995}
{Moles} M., {Marquez} I., {Perez} E., 1995, \apj, 438, 604

\bibitem[{{Park} \& {Choi}(2009)}]{Park2009}
{Park} C., {Choi} Y.~Y., 2009, \apj, 691, 1828

\bibitem[{P\'erez \& Granger(2007)}]{IPython}
P\'erez F., Granger B.~E., 2007, Computing in Science and Engineering, 9, 21

\bibitem[{{Petrosian}(1982)}]{Petrosian1982}
{Petrosian} A.~R., 1982, Astrofizika, 18, 548

\bibitem[{{Reichard} et~al.(2009){Reichard}, {Heckman}, {Rudnick},
  {Brinchmann}, {Kauffmann} \& {Wild}}]{Reichard2009}
{Reichard} T.~A., {Heckman} T.~M., {Rudnick} G., {Brinchmann} J., {Kauffmann}
  G., {Wild} V., 2009, \apj, 691, 1005

\bibitem[{{Rogers} et~al.(2009){Rogers}, {Ferreras}, {Kaviraj}, {Pasquali} \&
  {Sarzi}}]{Rogers2009}
{Rogers} B., {Ferreras} I., {Kaviraj} S., {Pasquali} A., {Sarzi} M., 2009,
  \mnras, 399, 2172

\bibitem[{{Sabater} et~al.(2012){Sabater}, {Verdes-Montenegro}, {Leon}, {Best}
  \& {Sulentic}}]{Sabater2012}
{Sabater} J., {Verdes-Montenegro} L., {Leon} S., {Best} P., {Sulentic} J.,
  2012, \aap, 545, A15

\bibitem[{{Sabater} et~al.(2013){Sabater}, {Best} \&
  {Argudo-Fern{\'a}ndez}}]{Sabater2013}
{Sabater} J., {Best} P.~N., {Argudo-Fern{\'a}ndez} M., 2013, \mnras, 430, 638

\bibitem[{{Schawinski} et~al.(2010){Schawinski}, {Dowlin}, {Thomas}, {Urry} \&
  {Edmondson}}]{Schawinski2010}
{Schawinski} K., {Dowlin} N., {Thomas} D., {Urry} C.~M., {Edmondson} E., 2010,
  \apjl, 714, L108

\bibitem[{{Silverman} et~al.(2009)}]{Silverman2009}
{Silverman} J.~D. et~al., 2009, \apj, 695, 171

\bibitem[{{Strateva} et~al.(2001)}]{Strateva2001}
{Strateva} I. et~al., 2001, \aj, 122, 1861

\bibitem[{{Strauss} et~al.(2002)}]{Strauss2002}
{Strauss} M.~A. et~al., 2002, \aj, 124, 1810

\bibitem[{{Tago} et~al.(2010){Tago}, {Saar}, {Tempel}, {Einasto}, {Einasto},
  {Nurmi} \& {Hein{\"a}m{\"a}ki}}]{Tago2010}
{Tago} E., {Saar} E., {Tempel} E., {Einasto} J., {Einasto} M., {Nurmi} P.,
  {Hein{\"a}m{\"a}ki} P., 2010, \aap, 514, A102

\bibitem[{{Tasse} et~al.(2011){Tasse}, {R{\"o}ttgering} \& {Best}}]{Tasse2011}
{Tasse} C., {R{\"o}ttgering} H., {Best} P.~N., 2011, \aap, 525, A127+

\bibitem[{{Taylor}(2005)}]{TOPCAT}
{Taylor} M.~B., 2005, in P.~{Shopbell}, M.~{Britton}, R.~{Ebert}, eds,
  Astronomical Data Analysis Software and Systems XIV. Astronomical Society of
  the Pacific Conference Series, Vol. 347, p.~29

\bibitem[{Walt et~al.(2011)Walt, Colbert \& Varoquaux}]{numpy}
Walt S.~v.~d., Colbert S.~C., Varoquaux G., 2011, Computing in Science \&
  Engineering, 13, 22

\bibitem[{{Wild} et~al.(2010){Wild}, {Heckman} \& {Charlot}}]{Wild2010}
{Wild} V., {Heckman} T., {Charlot} S., 2010, \mnras, 405, 933

\bibitem[{Woolf(1955)}]{woolf1955}
Woolf B., 1955, Annals of Human Genetics, 19, 251

\end{thebibliography}

\label{lastpage}

\end{document}